\documentclass[twoside]{article} 
\pdfoutput=1
\usepackage{indentfirst}
\usepackage{graphicx}
\usepackage[numbers]{natbib}
\usepackage{caption}
\usepackage{amsmath}
\usepackage{amsfonts}
\usepackage{amssymb}
\usepackage{mathrsfs}
\usepackage{bm}
\usepackage[normalem]{ulem}
\usepackage{epsfig,amssymb} 
\usepackage{multicol} 
\usepackage{CJK} 
\usepackage[dvipsnames,usenames]{color}
\usepackage{psfig}
\usepackage{hyperref}
\usepackage{epstopdf}
\input cyracc.def
\input amssymb.sty

\makeatletter
\newcommand\ackname{Acknowledgements}
\if@titlepage
  \newenvironment{acknowledgements}{%
      \titlepage
      \null\vfil
      \@beginparpenalty\@lowpenalty
      \begin{center}%
        \bfseries \ackname
        \@endparpenalty\@M
      \end{center}}%
     {\par\vfil\null\endtitlepage}
\else
  
\fi
\renewcommand\@biblabel[1]{$[{#1}]$}
\makeatother

\DeclareSymbolFont{lettersA}{U}{txmia}{m}{it}
\DeclareMathSymbol{\piup}{\mathord}{lettersA}{25}
\DeclareMathSymbol{\muup}{\mathord}{lettersA}{22}
\DeclareMathSymbol{\alphaup}{\mathord}{lettersA}{11}
\newfont{\yihao}{cmb10 at 19pt}
\newcommand{\fs}{\CJKfamily{fs}} 
\newcommand{\xiaosi}{\fontsize{11.5pt}{13.5pt}\selectfont} 
\renewcommand{\baselinestretch}{1.06} 
\catcode`@=11

\let\@oddfoot\@empty \let\@evenfoot\@empty
\def\@evenhead{\small\thepage\hfill {\small\it Fa-Bo Feng, Radio Jets and Galaxies as Cosmic String Probes, Front. Phys.}\hfill }
\def\@oddhead{\hfill{\small\it Fa-Bo Feng, Radio Jets and Galaxies as Cosmic String Probes, Front. Phys. }\hfill\small\thepage}

\def\hml{\end{multicols}\end{CJK*}\end{document}}
\newfont{\xbt}{cmb10 at 12pt}

\makeatletter
\long\def\@makecaption#1#2{%
\vskip\abovecaptionskip
\sbox\@tempboxa{#1\quad #2.}%
\ifdim \wd\@tempboxa >\hsize
  #1\quad #2.\par
\else
  \global \@minipagefalse
  \hb@xt@\hsize{\hfil\box\@tempboxa\hfil}%
\fi
\vskip\belowcaptionskip}
\makeatother

\begin{document}
\thispagestyle{empty}

\vspace*{-15mm} {\small Front. Phys.
\ \\
\small DOI 10.1007/s11467-011-0188-x\ }\\
\vspace*{-3.5mm} {{\hspace*{-1mm}\psfig{figure=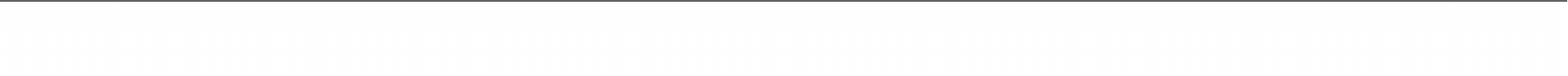}}}
\\[-1mm]

\textcolor{Orange}{{\yihao R}{\bf\xiaosi \!E\!S\!E\!A\!R\!C\!H}{\bf\xiaosi
\!A\!R\!T\!I\!C\!L\!E}}
 \vspace*{8mm}

\begin{center}
{\usefont{T1}{fradmcn}{m}{n}\yihao Radio Jets and Galaxies as Cosmic String Probes}
\footnotetext{\centerline{\hspace*{-10mm}\copyright Higher
Education Press and Springer-Verlag 2009}}
\vspace*{6mm}\\
{{\bf\small Fa-Bo Feng}
{\fs\xiaosi}$^{1}$}
\vspace*{5mm}

{\footnotesize\it $^{1}$Department of Astronomy, Nanjing University,
Nanjing 210093, P.R. China\\
E-mail: fengfabo@gmail.com}\vspace{1mm}

{\it \footnotesize \rm Revised January, 2011; accepted May 4, 2011}

\vspace{6mm}
\baselineskip 10pt
\renewcommand{\baselinestretch}{0.8}
\parbox[c]{145mm}
{\noindent{The lensing effect of a cosmic string is studied, and some new methods are
 proposed to detect the cosmic string. The technique for using jets as extended gravitational
  lensing probes was firstly explored by Kronberg. We use the "alignment-breaking parameter"
   $ \eta_G $ as a sensitive indicator of gravitational distortion by a wiggly cosmic string.
Then, we applied the non-constant deflection angle to jets, and $ \eta_G $ of a specific jet is just related to
the projected slope of the jet. At least three jets in the sample of Square Kilometer Array (SKA)
 would have significant signals ($ \eta_G >10^\circ $) if the wiggly infinite cosmic string
 existed.
The distortion of elliptical object is also studied and used to do a statistical research
 on directions of axes and ellipticities of galaxies. In the direction of the
 string, we find that galaxies appear to be more elliptical for an observer and the distribution of apparent
  ellipticity changes correspondingly. Ellipticity distribution of current SDSS spiral sample has the signal-to-noise
   ratio up to 8.48 which is large enough for astronomical observations. The future survey,
    such as Large Synoptic Survey Telescope (LSST) and Dark Energy Survey (DES), would weaken
     the requirement of special geometry in the data processing. As a result, all kinds of
     distributions, including ellipticity axis distribution, would serve as probes to detect wiggly strings in the near future. In brief,
     if a wiggly cosmic string existed, these signals would be convenient to be observed with
     the future weak lensing survey or other surveys in deep space. If there was no lensing signal
      in these distributions, it would give the upper limit of the abundance of infinite strings.
       }\vspace{2mm}

{\normalsize\bf Keywords\hspace{2.5mm}\rm gravitational lensing, jets and outflows,
cosmic string, galaxy }
\vspace{2mm}

{\normalsize\bf PACS~numbers\hspace{1.5mm}\rm  98.80.Cq,  98.62.-g, 98.62.Nx, 95.30.Sf
}}
\end{center}\normalsize

\baselineskip 12pt \renewcommand{\baselinestretch}{1.06}
\parindent=10.8pt \parskip=0mm \rm\vspace{1mm}

\begin{multicols}{2}\vspace*{-4mm}
\noindent
\chapter{\large
\usefont{T1}{fradmcn}{m}{n}\xbt 1\quad Introduction}
\vspace{2mm}

\noindent A cosmic string, as a kind of topological defect, is generated by symmetry breaking of
unified theories, such as fundamental string theory, brane inflation, M-theory, etc \cite{Ki76,ViSh00,Po04}. The detection of a cosmic
 string can be used to constrain different inflation models because strings were formed in the
 symmetry-breaking phase transition of some special inflation models, such as brane inflation
  \cite{Av07,Pe99,FeGaLiSoSo08}. Cosmic string gas may produce observable non-Gaussianity if
  the string scale is at TeV scale \cite{ChWaXuBr07}. Additionally, as a special kind of "dark matter",
   a cosmic string can produce cosmic ray positrons by the cosmic string cusp annihilation process,
   which might explain the positron excess of PAMELA and ATIC experiment \cite{BrCaXuZh09}. Moreover,
    a cosmic string may play a sub-dominant role in the formation of large structure of the universe
     to some extent \cite{Fa88,AcMa08}.

The gravitational properties of strings are studied with the weak-field approximation. According to
 the metric of a string, the light is deflected by an angle of $ 4\pi G U $. Thus, this gravitational
  effect could give rise to double images of background cosmic objects within the angle of order
   $ \delta\phi $ from the string \cite{Vi81,Go85,Wu90}. Their gravitational effect is measured by
    a dimensionless parameter $ G U /c^2 $, which is about $ 10^{-6} $ for GUT (grand unified theory)
     strings. Due to the gravitational effect caused by a cosmic string, many observations can be used
      to probe the properties of a cosmic string. The current result is $ G U /c^2 <0.7\times10^{-6} $,
       which is obtained by CMB constraints \cite{bevis08}. First LIGO search for gravitational wave
        bursts from cosmic strings also give nearly the same upper limit \cite{abbott09}. A cosmic
        string makes the space-time around it to be a conical space-time, resulting in the appearance
        of undistorted double images \cite{CoHu87,MaWeKi07,Wu89:Lensi}. Micro-lensing provided a way
         to detect lensing signal even when the image splitting is too small to resolve with astronomical
          measurements \cite{KuSiVa08}. With the development of the technology of weak lensing, the weak
           lensing effect produced by a cosmic string could also be a useful tool to constrain the
           properties of a cosmic string \cite{DyBr07,ThCoMa09}.

In this paper, we consider the weak lensing effect caused by a cosmic string, and polarized jets and
 elliptical objects are used as the background sources. Using the fact that the polarization angle
 of a background radio jet is not changed by gravitational distortion \cite{DySha92}, it is convenient
  to define an "alignment-breaking parameter" $\eta_G$ \cite{kronberg91}, which is sensitive to
   gravitational distortion. This method was used to analyze the polarization of the jet 3C 9 and
   give the redshift and properties of lens galaxies \cite{KrDyRo96}. Theoretical modeling of weakly
    lensed polarized sources is developed \cite{burns04} for the prediction of effects of gravitational
     lensing. As promising as it is, gravitational lensing is hampered by the lack of the knowledge of
      the intrinsic morphology of the sources being lensed. However, the polarization of the source
      could supply a useful probe to determine the intrinsic morphology of the source. Nevertheless,
      the lens employed by them are all spherical galaxies or elliptical ones. To our knowledge, this
      is the first investigation of weak lensing effect of a cosmic string with polarized sources as
       background objects. Once the background galaxy was lensed by a wiggly cosmic string, the
       ellipticity axis distribution (EAD), and the ellipticity distribution (ED) would be changed.
        Though there might be just several infinite straight strings \cite{Wu89:Lensi}, they would
         distort the light passing by it with any impact parameter. For a straight string, the deflection
          angle of an image is constant. Actually, there is no absolutely straight string. And the wiggly string would produce non-constant deflection angle which would result in the distortion of background sources\cite{DyBr07}. In this paper, we will consider different kinds of sources as background objects to probe the property of the wiggly cosmic string.

The paper is organized as follows: First, three basic concepts, alignment-breaking parameter, EAD and ED, are introduced and the deflection angle is deduced. Second, we use the "alignment-breaking-angle"
  to study lensing effect of a straight jet. EAD and ED of galaxies are also studied. Then, we put the
  formulas into simulations and provide the strategy to probe the signals produced by weak lensing effects
   of a cosmic string. In the part of conclusion, we propose some statistical methods to search for such
   signals in the future surveys.\vspace*{4mm}

\noindent
\chapter{\large\usefont{T1}{fradmcn}{m}{n}\xbt 2\quad Theory}
\vspace{1.8mm}

\section*{\normalsize\rm 2.1\quad Basic concepts}

\subsection*{\normalsize\it 2.1.1\quad Alignment-breaking parameter $ \eta_G $}
\begin{figure*}[ht]
\centering
\includegraphics[scale=0.3]{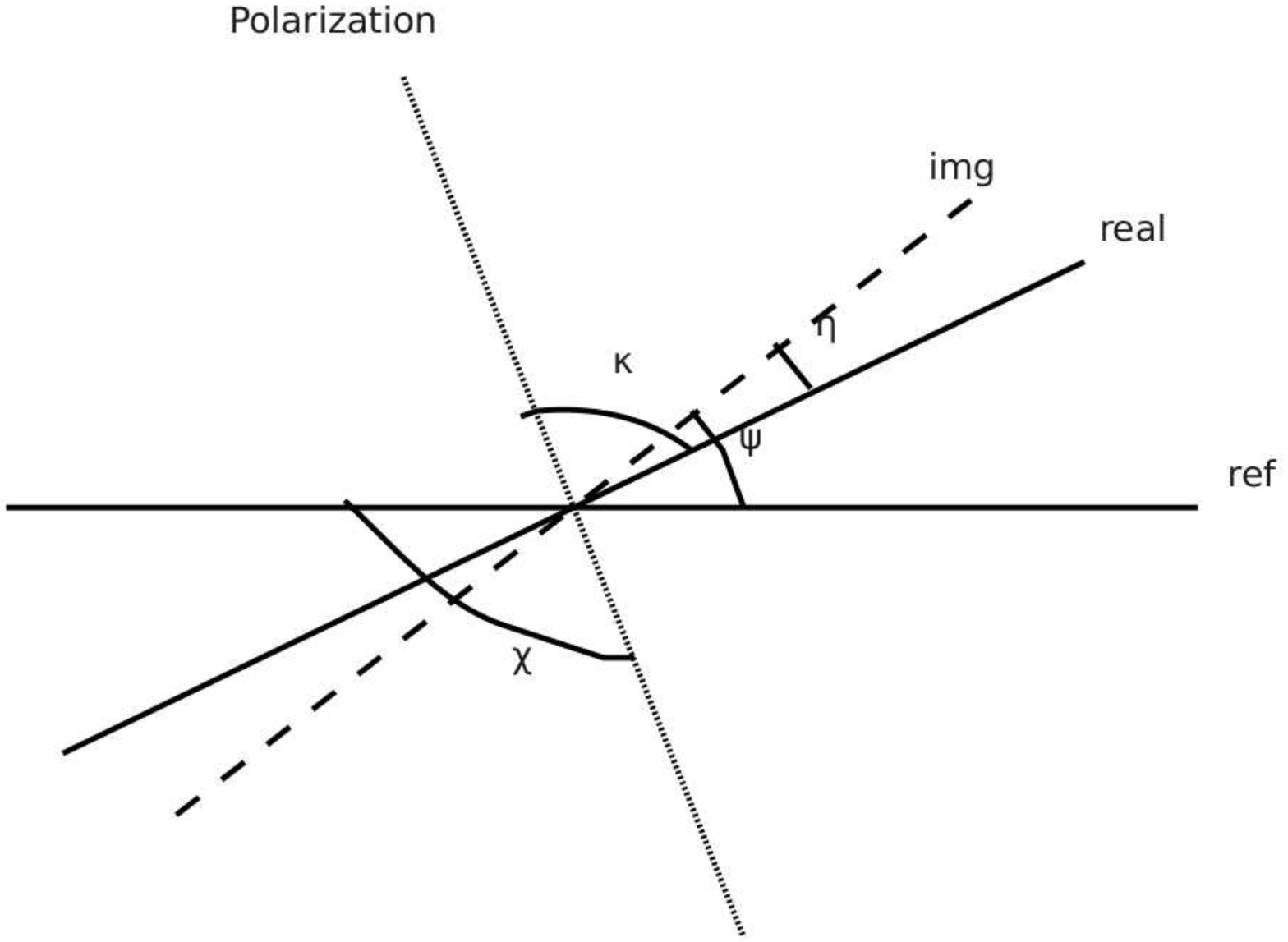}
\caption{The relation of the alignment-breaking parameter, $ \eta_G $, to other angles $ \kappa $, $ \psi $,
 $ \chi_0 $. The dotted (dashed) line represents polarization (image), which is the projection of the direction
  of polarization on the image plane }
\label{fig:etaG}
\end{figure*}

\vspace{1mm}
\label{etaG}
\noindent In order to measure the signals yielded by a cosmic string, we should at first define a parameter
 $ \eta_G $ \cite{kronberg91} as the angle between the direction of observed polarization and the tangential
  vector of a jet:
\begin{equation}
\eta_G (\vec{\theta})=\psi(\vec{\theta}) -\chi_0 (\vec{\theta})+\kappa(\vec{\theta})
\label{eq:para}
\end{equation}
where $ \psi(\vec{\theta}) $ is the angle of tangent vector relative to the reference line of the jet at the
 projected position $ \vec{\theta} $.
$ \chi_0$ is the polarization vector, and $ \kappa $ is some intrinsic deviation from a perfect jet
 (see Fig.\ref{fig:etaG} ).
Define $ \vec{e}_{img},\vec{e}_p,\vec{e}_{real},\vec{e}_{ref} $ as the unit tangent vector of the fiducial
line of the image, direction of polarization, fiducial line of the source, direction of reference line,
 and the vectors corresponding to $ \psi ,\chi_0,\kappa $ ar
\begin{align}
\vec{\kappa}=&\vec{e}_p -\vec{e}_{real} ,\nonumber\\
\vec{\chi}_0=&\vec{e}_p -\vec{e}_{ref} , \nonumber\\
\vec{\psi}=&\vec{e}_{img} -\vec{e}_{ref} ,\nonumber\\
\vec{\eta}_G=&\vec{e}_{img} -\vec{e}_{real} ,
\label{eq:etaillustration}
\end{align}
so $ \eta_G =\psi +\kappa -\chi_0 $.
$ \eta_G $ is a measure of the projection of shear onto the jet itself.\par
In equation (\ref{eq:etaillustration}), the polarization, $ \chi_0 $, tangent angle, $ \psi $, and the intrinsic deviation of the
 polarization from the tangent angle, $ \kappa_0 $, must be measured along the jet or lobe. However, the
  intrinsic polarization of the source, $ \chi_0 $, is coupled with the Faraday Rotation, which causes the
   polarization of a photon to rotate as it traverses a magneto-ionic medium. By measuring Faraday rotation
   at different wave lengths, one could eliminate the variation of $ \eta_G $ brought by Faraday rotation.
   The true value of rotation measure is needed if we want to rule out the constant part produced by Faraday
   rotation \cite{Burns02}. In addition, the value of  $ \kappa $ is a source-dependent angle that allows
   for a source-intrinsic variation of $ \eta_G $. Fortunately, the polarization of a jet is highly aligned
   with the local direction of the jet \cite{PeBrWi84,KiBiEk86,BrPeHe86}. Simulations of the intrinsic
   magnetic field of a radio jet are conducted by using MHD model of a jet \cite{PeBrWi84}. Constrains
   on $ \kappa $ would require the observation of a low-redshift sample of sources, and even then, $ \kappa $
   could only be understood in a statistical sense \cite{Burns02}. Therefore, with the mature theory of the
   magnetic field of jets, we could independently derive the value of $ \kappa(\vec{\theta}) $ and consequently
   use $ \eta_G $ to reveal the strength of the gravitational bending varies along the jet. That is, observing
    a jet and its polarization could confirm the value of $ \eta_G $, which is deduced from the gravitational
    effect of a cosmic string. Thus, $ \eta_G $ serves as another parameter to reveal the property of a cosmic
     string.

\vspace{1mm}

\subsection*{\normalsize\it 2.1.2\quad The ellipticity and ellipticity axis angle of the galaxy}\vspace{1mm}

\noindent
For a spiral or elliptical galaxy, the intrinsic shape might be simplified into a triaxial ellipsoid,
$ x^2/a^2+y^2/b^2+z^2/c^2=1 ~with~a\geqslant b\geqslant c$. The galaxies can be divided into three kinds:
 oblate galaxy (a=b), prolate galaxy (b=c), and triaxial galaxy ($ a\neq b\neq c $). Define
 $ \gamma\equiv c/a $ as the thickness of a galaxy and $ \varepsilon\equiv 1-b/a $ as the
 intrinsic ellipticity, we can convert the intrinsic coordinate system to the observer's
  coordinate system and compute the resulting apparent axis ratio $ \epsilon \equiv b_{ob} /a_{ob} $
   and the ellipticity axis angle (EAA) $ \Psi $ of a triaxial galaxy \cite{Bi85}:
\begin{align}
&\epsilon=\left[\displaystyle\frac{X+Z-\sqrt{(X-Z)^2+Y^2}}{X+Z+\sqrt{(X-Z)^2+Y^2}}\right]^{1/2} ,\label{eq:epsilon}\\
&\Psi=\displaystyle\frac{1}{2}\arctan{\left(\displaystyle\frac{Y}{X-Z}\right)},\label{eq:psi}
\end{align}
where
\begin{eqnarray*}
&X=[1-\varepsilon(2-\varepsilon)\sin^2{\varphi}]\cos^2{\theta}+
\gamma^2\sin^2{\theta},\\
&Y=\varepsilon(2-\varepsilon)\cos{\theta}\sin{2\varphi},\\
&Z=1-\varepsilon(2-\varepsilon)\cos^2{\varphi},
\end{eqnarray*}
and $ {\theta,\phi} $ is the spherical coordinates of the line of sight.
The photometric analysis of Sloan Digital Sky Survey (SDSS) provide a model-free
measures of the axis ratio by the technique of adaptive moments. This method guarantees
 the accuracy of the values of $ a_{ob} , b_{ob} $, which are the long and short semi-major
  axis lengths.
By fitting models to the shape measurement of SDSS Data Release 1 (DR1), the observed apparent
 axis ratios could be modeled by adopting a Gaussian distribution of $ \gamma $ and a log-normal
 distribution of $ \varepsilon $ \cite{Ry04}:
\begin{eqnarray}
&f(\gamma)\propto exp\left[-\displaystyle\frac{(\gamma-\mu_\gamma)^2}{2\sigma_\gamma^2}\right]^{1/2}~~~~~0\leqslant\gamma\leqslant1,\\
&f(\varepsilon)\propto \displaystyle\frac{1}{\varepsilon} \displaystyle\exp\left[-\frac{(\ln\varepsilon-\mu)^2}{2\sigma^2}\right]~~~~~\ln{\varepsilon}<0.
\label{eq:EDEAD}
\end{eqnarray}
With four parameters $ \mu_\gamma,\sigma_\gamma,\mu,\sigma $ estimated from the data,
we could describe the shapes of elliptical
galaxies and spiral galaxies with the distribution of apparent axis ratio.\vspace{1.5mm}

\section*{\normalsize\rm 2.2\quad Deflection angle by a wiggly cosmic string}
\begin{figure*}
\begin{center}
\includegraphics[scale=0.5,angle=90]{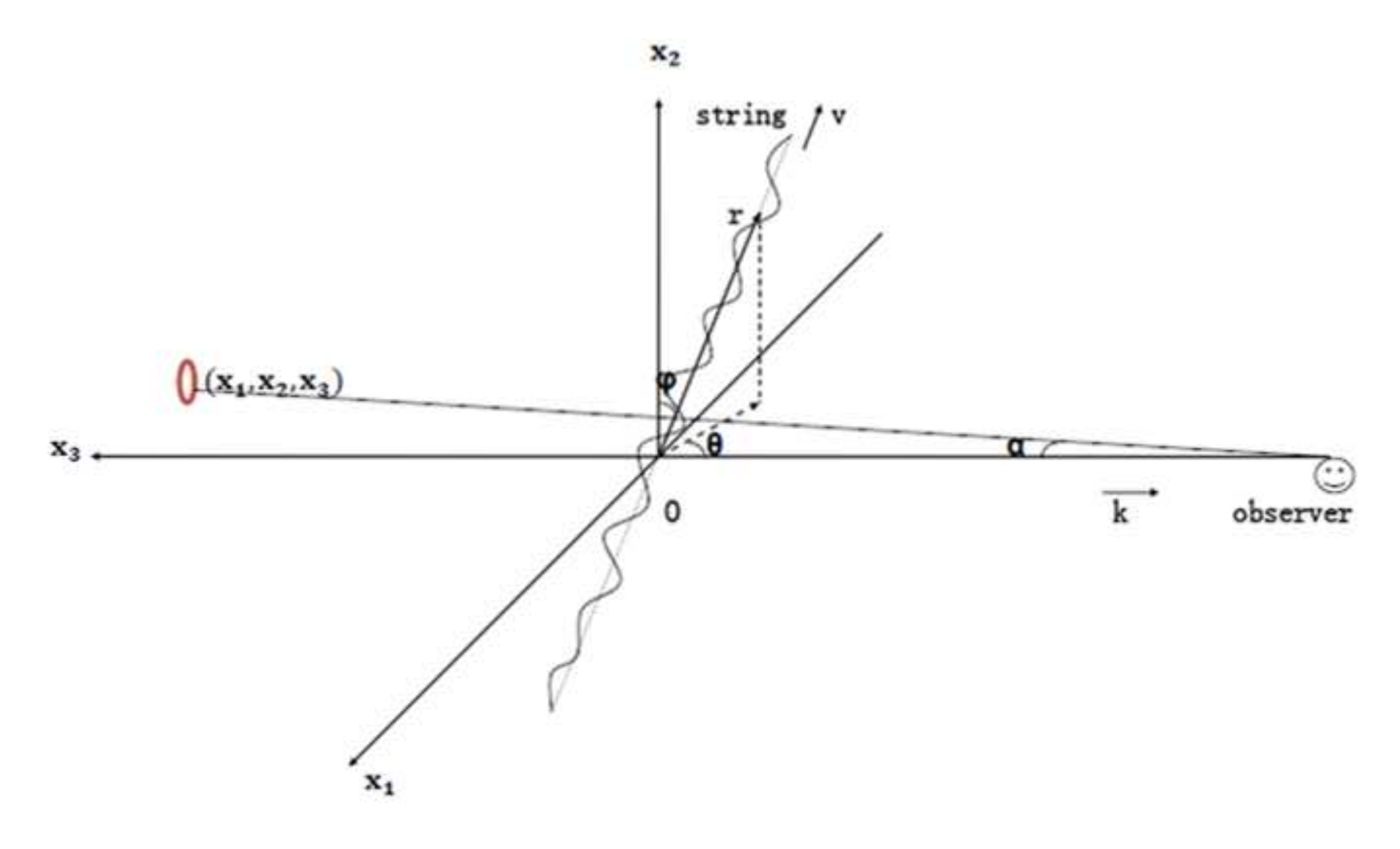}
\caption{The wiggly string leads to some distortion of the straight jet\label{fig:string}}
\end{center}
\end{figure*}

\begin{figure*}
\begin{center}
\includegraphics[width=50mm]{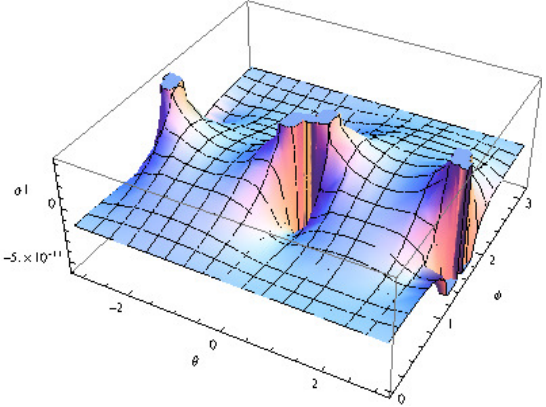}
\includegraphics[width=50mm]{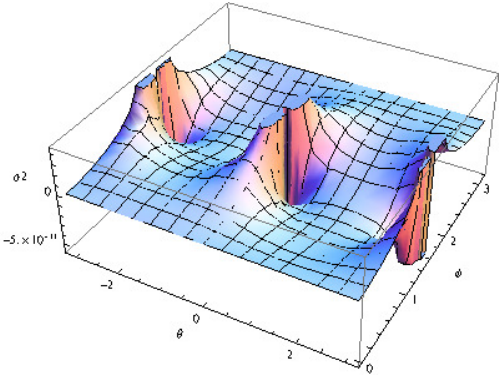}
\includegraphics[width=50mm]{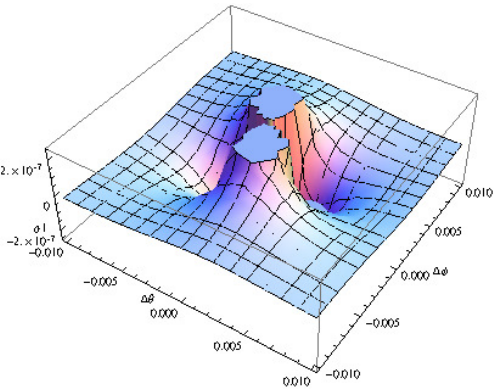}
\caption{The left graph shows the 3D plot of $\alpha1$ and the middle one shows that of $\alpha2$. Notice that the two graphs are all plotted in the limitation where $\{\theta,\phi\}$ is not close to $\{0,\pi/2\}$. The right one illustrate the value of $\vec{\alpha}$ when $\{\theta,\phi\}\rightarrow\{0,\pi/2\}$, where $\{\theta,\phi\}\equiv\{\Delta\theta,\pi/2+\Delta\phi\}$.}
\label{fig:alpha12}
\end{center}
\end{figure*}

\vspace{2.3mm}
\noindent We quote the result of ref.\cite{uzan00} for the deflection of a light ray by a static wiggly cosmic string,
\begin{equation}
\vec{\alpha}=-\vec{\nabla}_\bot\int_{-D_L}^{D_S -D_L}\Phi(x_1,x_2,x_3)d x_3 ,
\label{eq:deflection}
\end{equation}
where $D_S$ and $D_L$ are separately the distance from the observer to the source and to the cosmic string. Notice that the analytic expression of the deflection angle does not include the constant deflection angle which is of no use for distorted background objects. The deflecting potential is given by
\begin{equation}
\Phi(x_1,x_2,x_3)=2G\int\frac{\mathscr{F}_{\mu\nu}k^\mu k^\nu}{|\vec{x}-\vec{r}|}d\ell .
\label{eq:phi1}
\end{equation}
Here G is the gravitational constant, $\vec{x}$ is a reference point, $k^\mu\equiv dx^\mu/d\lambda$ is the tangent vector to a geodesic $g_o$ which is chosen as reference (see Fig.\ref{fig:string}). Also, $\vec{r}(\ell)$ is a parametrization of the string, in terms of which
\begin{equation}
\mathscr{F}_{\mu\nu}=T_{\mu\nu}-\frac{1}{2}\eta_{\mu\nu}T_\lambda^\lambda ,
\label{eq:fmunu}
\end{equation}
where $ T_{\mu\nu}$ is the stress-energy tensor and $\eta_{\mu\nu}$ is the Minkowski metric, i.e. $\eta_{\mu\nu}=diag(-~+~+~+)$.
Generally speaking, the stress-energy tensor of a cosmic string is of the form
\begin{equation}
T^{\mu\nu}=U u^\mu u^\nu-T v^\mu v^\nu,
\label{eq:tmunu}
\end{equation}
where U is the energy per unit length, T is the tension which doesn't equal to U for a wiggly string, $u^\mu$ is the 4-velocity of the observer ($u^\mu$ is also a timelike vector, i.e. $u^\mu u_\mu =-1$) and $v^\mu$ is a spacelike ($v^\mu v^\nu=1$) unit vector tangent to the string word sheet. Let us consider a tilted static cosmic string aligned as Fig.\ref{fig:string} illustrated, and the two unit vector are
\begin{equation}
\begin{aligned}
u^\mu=&(1,0,0,0),\\
v^\mu=&(0,-\sin{\theta}\sin{\phi},\cos{\phi},-\cos{\theta}\sin{\phi}).
\label{eq:uv}
\end{aligned}
\end{equation}
We take the tangent vector of the reference geodesic to be $k^\mu=(1,0,1,0)$. From (\ref{eq:fmunu})-(\ref{eq:uv}), we deduce that
\begin{equation}
\begin{aligned}
\mathscr{F}=&\mathscr{F}_{\mu\nu}k^\mu k^\nu\\
=&(T_{\mu\nu}-\frac{1}{2}T^\lambda_\lambda\eta_{\mu\nu})k^\mu k^\nu\\
=&U-T\sin^2{\phi}\cos^2{\theta}.
\end{aligned}
\end{equation}
If we take $\theta=0$, then $\mathscr{F}=U-T\sin^2{\phi}$, which is the result of ref. \cite{uzan00}. When $U(\ell)$ and $T(\ell)$ are constants, the deflecting potential, after integration over $x_3$, reduces to
\begin{equation}
\begin{aligned}
\Phi(x_1,x_2,x_3)=&-4G(U-T\sin^2{\phi}\cos^2{\theta})\ln(\vec{x}^2-\\
&(-x_1 \sin{\theta}\sin{\phi}+x_2\cos{\phi}-x_3\cos{\theta}\sin{\phi})^2).
\label{eq:phi2}
\end{aligned}
\end{equation}
In the limit where $|x_3|\gg |x_1|,|x_2|$ and under the condition that $\{\theta,\phi\}$ is not very close to $\{0,\pi/2\}$, the deflection angle, i.e. $\vec{\alpha}(x_1,x_2)=\{\alpha1,\alpha2\}$, is deduced from (\ref{eq:deflection}). However, the form of it is complicated, so we have plotted the 3D graph of $\vec{\alpha}(\theta,\phi)$ (see Fig.\ref{fig:alpha12}) after taking $G U=1.1\times10^{-6}$,$G T=0.9\times10^{-6}.$ \footnote[1]{Here, G, U, T are all in nature unit. We adopt the value of GUT string which is produced during the GUT transition, $E=10^15 GeV$.}Reasonably, we take $p\equiv D_L/D_S=1/2$ and $,|x_1|/D_S=|x_2|/D_S=10^{-6}$.

Illustrated by Fig.\ref{fig:alpha12}, the deflection angles, $\alpha1$ and $\alpha2$, are sensitive to the change of $\{\theta,\phi\}$ especially when they come close to $\{0,\pi/2\}$. Additionally, the signs of the two deflection angles are almost opposite. However, this result is limited by its assumption of the value range of $\{\theta,\phi\}$. Therefore, it is necessary to explore the form of deflection angle in the limit where $\{\theta,\phi\}\rightarrow\{0,\pi/2\}$ in order to enhance the value of deflection angle.
After definition of $\{\theta,\phi\}\equiv\{\Delta\theta,\pi/2+\Delta\phi\}$ and the limitation where $\Delta\theta\rightarrow0$ and $\Delta\phi\rightarrow0$, the form of defecting potential of (\ref{eq:phi1}) reduces to
\begin{equation}
\Phi(x_1,x_2,x_3)=4G(U-T)\ln[(x_3 \Delta\phi + x_2)^2 + (x_3 \Delta\theta + x_1)^2],
\end{equation}
where $\Delta\theta$ and $\Delta\phi$ should be larger than $10^{-6}$ which makes $x_3 \Delta\theta$ and $x_3 \Delta\phi$ much larger than $x_1$ and $x_2$. Then, we recall the form of deflection angle in (\ref{eq:deflection}), which reduces to
\begin{equation}
\begin{aligned}
\alpha1=&\alpha2\\
&=-\frac{4 G (T-U) \left(2 x_2\Delta\theta \Delta\phi +x_1\left(\Delta\theta^2-\Delta\phi^2\right)\right) D_S}{\left(\Delta\theta ^2+\Delta\phi ^2\right)^2 D_L \left(D_L-D_S\right)}.
\label{eq:alpha12}
\end{aligned}
\end{equation}
From the equations above, we find that the deflection angle is proportional to $(\Delta\theta ^2+\Delta\phi ^2)^{-1}$ approximately.
To compare the deflection angle deduced from (\ref{eq:phi1}) with that deduced from (\ref{eq:phi2}), we have plotted the graphs of $\vec{\alpha}$ after adopting the above approximation(see Fig.\ref{fig:alpha12}). As a result, we find that the general shapes of them are similar with those of Fig.\ref{fig:alpha12}. However, the sign of the two deflection angle is not opposite now. Because we take $x_1/D_S =x_2/D_S =10^{-6}$ and the deflection angle is up to $10^{-7}$, the distortions of the background objects are large enough to be observed. The statistics of background jets and galaxies would reveal the secrets of the wiggly cosmic string.
\vspace*{4mm}

\noindent
\chapter{\large
\usefont{T1}{fradmcn}{m}{n}\xbt 3\quad Simulations}\vspace{0.5mm}

\section*{\normalsize\rm 3.1\quad Radio jets}
\vspace{2.3mm}
\noindent Considering the statistical errors and systematic errors, we can only extract the signal of $ \eta_G >10^\circ $ (i.e., $ \eta_G >0.175~ rad $) \citep{Burns02} from the polarization analyzing of radio jets. So we should take the approximation that $\Delta\theta,~\Delta\phi\rightarrow0$ to yield large distortion angle.

\begin{figure*}
\begin{center}
\includegraphics[width=70mm]{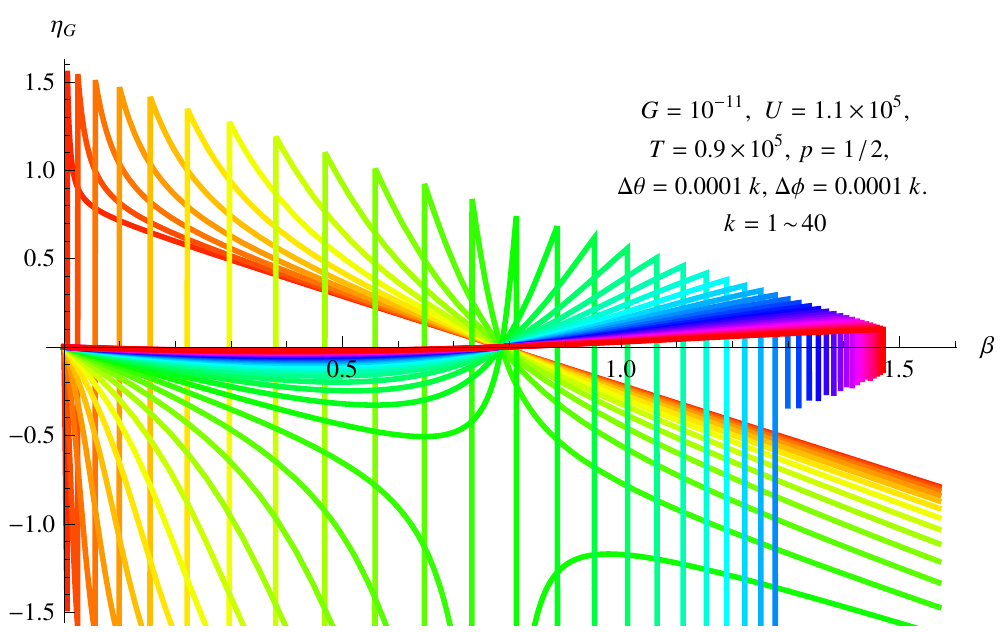}
\includegraphics[width=70mm]{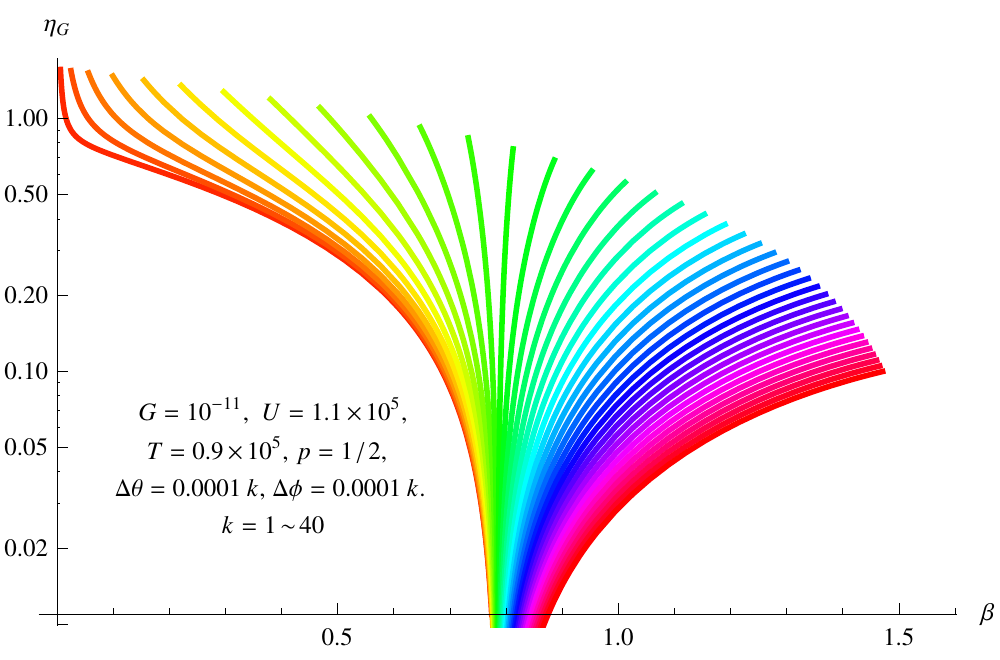}
\caption{The left graph shows the 2D plot of $\eta_G$ and the right one shows the log plot of $\eta_G$ as a function of $\beta$. Notice that the two graphs are all plotted by taking integer k from 0 to 40 colored from red to blue}
\label{fig:etaGseries}
\end{center}
\end{figure*}

\begin{figure*}
\centering
\includegraphics[width=120mm]{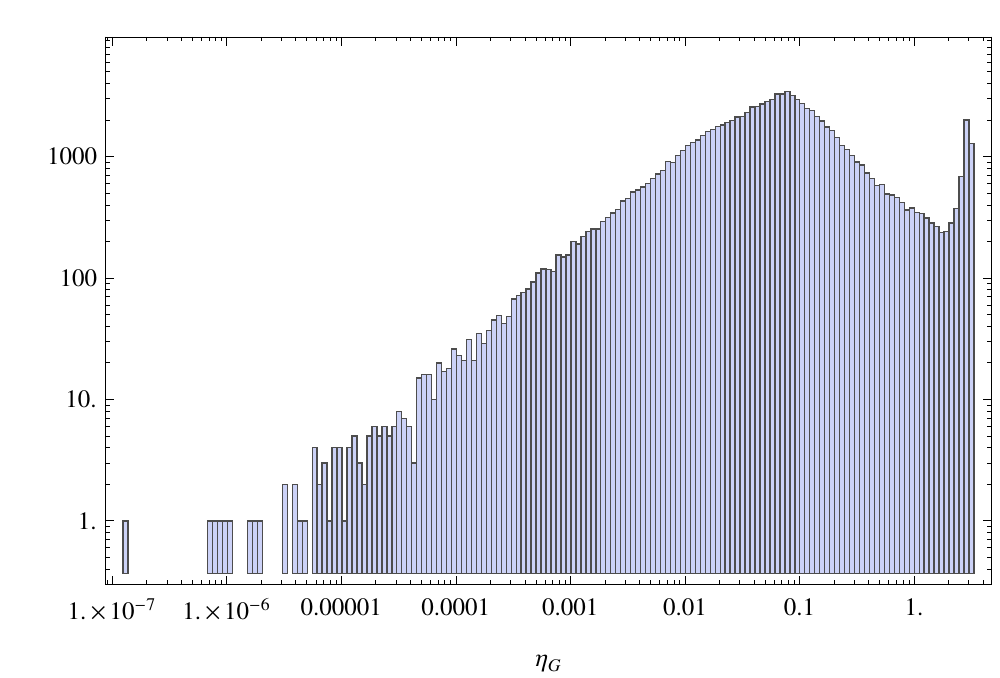}
\caption{The distribution of $ \eta_G $ is illustrated by this graph in which log-log
 coordinates make the graph more appropriate. There are about 100,000 points for statistics}
 \label{fig:EtaGhistogram}
\end{figure*}

First, we have parameterized the configuration of the radio jet as follows:
\begin{equation}
\begin{aligned}
x_1=&r\cos\beta +x_{1c},\\
x_2=&r\sin\beta +x_{2c},
\end{aligned}
\end{equation}
where r is the distance from reference point to the center of the jet, and $\beta$ is the angle between its vector and the axis of $x_1$.
The distortion could be deduced as $x'_1=x_1+D_S\alpha1 $ and $x'_2=x_2+D_S\alpha2 $. The corresponding $\eta_G$ reduces to
\begin{equation}
\begin{aligned}
\eta_G\equiv&\Delta\beta\\
=&\arctan[(4G(T-U)(\Delta\theta ^2-\Delta\phi ^2)+(8 G (T-U) \Delta\theta \Delta\phi\\
&+p(\Delta \theta ^2+\Delta\phi ^2)^2-p^2(\Delta\theta ^2+\Delta\phi ^2)^2) \tan\beta)\\
&/(4 G (T-U) (\Delta\theta^2-\Delta\phi^2)-(-1+p)p(\Delta\theta^2+\Delta\phi^2)^2\\
&+8G(T-U)\Delta\theta \Delta\phi \tan\beta)]-\beta.
\end{aligned}
\end{equation}
From the equation above, we find that $\eta_G$ is independent with the coordinates of the points on the jet, i.e. $\eta_G$ is constant for a specific straight jet. However, $\eta_G$ is sensitive to the inclined angle $\beta$. The shape of $\eta_G$ is different with the sinusoidal $\eta_G$ shape yielded by spherical lens \cite{kronberg91}. However, cosmological birefringence could also result in a rotation of the polarization of photons, which could yield a nonzero $ \eta_G $ \citep{Kam10}. Though this kind of effect may appear similar with that caused by a cosmic string, they have different features. Birefringence is a cosmological effect that could be detected in the whole universe with nearly the same order of magnitude. That is, if $ \eta_G \neq 0 $ for one jet, it would also be founded in other jets. However, the signature given by a wiggly cosmic string is local. Of course, other sources, such as gravitational waves and rotational black holes may also yield nonzero $ \eta_G $, but their shapes are not constant for different points on the jet\citep{DySha92,Far08}. Additionally, with current resolution of the radio telescope, such as SKA, the main errors come from the intrinsic noise of the polarization $ \kappa $, the error of choosing the anchor points, and the residual noise after extracting the effect of Faraday Rotation measurement (RM). Considering these factors, we can only extract the signal of $ \eta_G >10^\circ $ \citep{Burns02} from the polarization analyzing of radio jets. Thus, it is important to yield large enough signals by carefully tune the parameters. With current observation, the properties of the cosmic string are parameterized by $GU=1.1\times10^{-6}$ and $GT=0.9\times10^{-6}$; and the $D_L$ to $D_S$ ratio $p$ is taken to be 1/2. We could plot the 2D graph series with different ranges of $\Delta\theta$ and $\Delta\phi$ (see Fig. \ref{fig:etaGseries}). From these two graphs, we find that there would be signals large enough to be observed when $\Delta\theta=\Delta\phi\simeq0.001$. Notice that the value of the signal is always proportionate to $G(U-T)$. And $\eta_G$ is sensitive to the angle of inclination $\beta$, which also give us the information of the configuration of the jet.
In order to get a general idea of the probability for the detection of a cosmic string through analyzing $ \eta_G $, we put the distribution of $ \eta_G $ into programming and draw the graph by Mathematica. $ \Delta\theta,\Delta\alpha $ are randomly chosen in the range of $[-0.005,0.005]$. Larger interval result in a little proportion of useful points though the sample is large at the interval; and a smaller interval would lead to a very small sample which is not appropriate for statistics. The number of background objects was taken to be 100,000, which calls for 100,000 times of random choices of these parameters in their intervals. The parameters of cosmic string are set to be the same as the above ones, and the lower limit of rotation angle or $\eta_G$ is taken to be $ \eta_G >10^\circ $ (i.e., $ \eta_G >0.175~ rad $). After that, the final result is revealed by Fig.\ref{fig:EtaGhistogram}.

In this histogram, we find about 1/5 jets would yield rotation angles larger than $10^\circ$ when $ \Delta\theta$ and $\Delta\alpha$ are at the interval of $[-0.005,0.005]$. Notice that 100,000 jets in such a small interval do not accord with reality. We adopt such a large sample is just for getting the probability of producing large signals. Though, the population in such an interval is small currently, SKA may change the situation. It would carry out an All-Sky RM Survey, which could yield RMs for approximately $ 2 \times 10^4 $ pulsars and $ 2\times10^7 $ compact polarized extragalactic sources in about a year of observing time \citep{GaeBecFer04}. In this All-Sky RM survey, we may get useful information of the intrinsic polarization angle of pulsars who have jets and lobs. To be on the safe side, we consider $2\times10^6$ radio objects as an ideal sample. If the wiggly cosmic string or the string web exists, the population of the useful jets in the sub-sample of the right area is $1/5\times(0.01)^2\times2\times10^6/(4\pi)\simeq 3$. And in other areas of the sky where $\Delta\theta$ and $\Delta\phi$ larger than 0.01, there would be nearly no observably signal. Notice that these jets are characterized by constant rotational angle $ \eta_G $.

In sum, the value of $ \eta_G $ is determined by the relative position of the jet with respective to the cosmic string and the inclined angle of the string. The position where $ \eta_G $ would be enhanced is in the direction of the string; and the larger the inclined angle is, the larger the deflection angle is. Additionally, constant $ \eta_G $ resulted from the straight jet is easily to be extracted from other noises. If $\eta_G$ is irregular, it would give us information about the configuration of the jet. In the near future, with the improvement of the accuracy of RM measurement and the maturation of theoretical model of the intrinsic magnetic field in the jet, $ \eta_G $ would show its strong power of detecting the wiggly string.\vspace{0.5mm}

\section*{\normalsize\rm 3.2\quad Galaxies}
\vspace{2.3mm}
\noindent In order to study the cosmic string with current large scale survey, it
is best to use galaxies as background sources. And the weak gravitational effect from a cosmic string should be analyzed with statistical method. Additionally, the method used here is a little different from that used in the statistics for the cosmic shear. The following method only abstracts the local information, and the study of cosmic shear aims at getting statistical information. The ellipticity axis distribution (EAD) of circular 2D sources was used to examine the shear signature of nearly circular sources by Dyda and Brandenberger \citep{DyBr07}. In our paper, EAD might also be a good parameter to describe the distortion of the EAA of an arbitrary galaxy in the data processing of the future survey. Additionally, we used the ellipticity distribution (ED) as a probe for the cosmic string detection. Rather than using the circular object as the source, we extended the shape of the source to be a 3D triaxial object. Then we project the 3D image onto the surface perpendicular to the viewer. Points on the surface of the elliptical object have coordinates: $ (x_{1c}+x_1,x_{2c} +x_2,x_{3c} +x_3) $.

In the following formulas, we use $ \epsilon \equiv b_{ob}/ a_{ob} $ to be the apparent axis ratio. And the luminosity density
 $ l $ is
\begin{equation}
l=l(a_v),~~~~~~
where~~
a_v\equiv\sqrt{x_0^2+\frac{y_0^2}{\zeta^2}+\frac{z_0^2}{\xi^2}}.
\end{equation}
The above formula is used by Stark and Binney to calculate the apparent axis ratio
\citep{STARK,Bi85}. However, we adopted a different transformation from the two-parameter
 transformation by them. They just use $ \theta,\phi $ to fix the system of the observer
 by making x' axis lie in the O-xy plane \citep{Bi85}. we adopted $ \alpha,\beta,\gamma $
 to describe the system of the observer. The system of $\{x_0,y_0,z_0\}$ is a system in which the axes of $ x_0,y_0,z_0 $ are aligned with the principal body axes of the ellipsoid. After transformation from this system to the coordinate system of the observer, their coordinates would reduce to
\begin{equation*}
\begin{aligned}
x_0=&x_1 \cos{\beta} \cos{\gamma} - x_3 \sin{\beta} +
 x_2 \cos{\beta} \sin{\gamma},\\
y_0=&x_3\cos{\beta} \sin{\alpha} +
 x_1(\cos{\gamma} \sin{\alpha} \sin{\beta} -
    \cos{\alpha} \sin{\gamma}) \\
 +&x_2 (\cos{\alpha} \cos{\gamma}+  \sin{\alpha} \sin{\beta} \sin{\gamma}),\\
z_0=&x_3 \cos{\alpha} \cos{\beta} +
 x_1 (\cos{\alpha} \cos{\gamma} \sin{\beta}+ \sin{\alpha} \sin{\gamma})\\
  +& x_2 (-\cos{\gamma} \sin{\alpha}+\cos{\alpha} \sin{\beta} \sin{\gamma}).
\end{aligned}
\end{equation*}
With the above transformation, we could get the elliptical radius, $ a_v(x_1,x_2,x_3) $, in the
 coordinate system of observer. With the method explored by Stark and Binney \citep{STARK,Bi85}, the apparent axis ratio of
    the ellipsoid on the sky will be easily deduced. Due to the complexity of formulas, the
    calculations are left for the computer to do. At last, the change of the ellipticity
     distribution (ED) is deduced from the distortion of $\epsilon $,
$ \Delta\epsilon=\widetilde{\epsilon}(\alpha,\beta,\gamma,\zeta,\xi,\theta,\phi)-\epsilon(\alpha,\beta,\gamma,\zeta,\xi) $.
Therefore, both ED and EAD are determined by distributions of these parameters. Among these parameters,
the distributions of $ \alpha,\beta,\gamma $ are determined by the random distribution of points on an unit sphere; $ \theta$ and $\phi $ are replaced by $\Delta\theta$,$\Delta\phi$, which obey the flat distribution; and $ \zeta,\xi $ obey Normal and Log-Normal distributions respectively.

It is obvious that $ \Delta \epsilon $ and $ \eta_G $ are determined by the value of $\vec\alpha$. That is, inserting (\ref{eq:alpha12}) in $a_v(x_1+D_S \alpha_1,x_2+D_S \alpha_2,x_3)$, we deduce the distorted apparent axis ratio (\ref{eq:epsilon}) and EAA (\ref{eq:psi}). $ \Delta \epsilon $ would also contribute to the cosmic shear if a wiggly cosmic string existed. The shear signal yielded by the cosmic string is subtle when $\theta\nrightarrow0$ and $\phi\nrightarrow\pi/2$. However, it may contribute a little fraction to the cosmic shear and result in systematic errors. The shear field of the wiggly cosmic string should also be studied for this reason \cite{ThCoMa09,Mack07}. If $\Delta\theta,~\Delta\phi\rightarrow0$, $ \Delta \epsilon $ would be proportional to $ (\Delta\theta^2+\Delta\phi^2)^{-1}$ and increase even to the order of 0.1. It will be interesting to find these statistical signals produced by the network of strings in the future weak lensing survey.

 Similarly, with the statistical method used in the last section, we could get the variation of signal-to-noise ratio (SNR) with the length of the interval of $\Delta\theta$ and $\Delta\phi$, i.e. $[-\delta/2,\delta/2]\equiv[\Delta\theta_{min},\Delta\theta_{max}]\equiv[-0.001k,0.001k]$. It is necessary to study SNRs of different modes (specified by integer k) of sky survey to know the feasibility of this method. Because the signal of ED/EAD and the noise of the sample in the area of $ \delta^2 $ increase with $\delta^{-2}$ simultaneously, we should find the optimal $\delta$, i.e. $ \delta_{opt}$, to make SNR highest. The specific processes are as follows:

First, considering ED/EAD without a string (original ED/EAD), we produced a realization of ellipsoids according to the distributions explored by Padilla and Strauss \citep{PaSt08}. Here, we assume the empirical ED is ideal because the fitting error is less than the statistical error especially in the interval of apparent axis ratio, $ \epsilon\in[0,0.1] $. The population of the sample shall be calculated according to the surface number density of the survey sample. For example, there are about 90 targets per square degree in the main galaxy sample of SDSS DR6. Thus, a realization of ED/EAD would call for about $ 90~\delta^2\times180^2/\pi^2 $ times of generations of random points. The parameters included in ED/EAD are $ \alpha,\beta,\gamma,\zeta,\xi$. $ \alpha,\beta,\gamma$ are determined by the random distributed points on an unit sphere. $ \gamma,\zeta $ are chosen according to the empirical ED. $ \Delta\theta $ and $\Delta\phi$ are randomly chosen in the interval of $[-0.001k,0.001k]$.

Second, by repeating the first step 100 times, we computed the mean and standard deviation of the expected number of original ellipsoids in each of the 10 bins. Additionally, the mean of distorted ellipsoids is also gotten simultaneously. The standard deviation is the function of $ \delta$. By the way, the reason for setting 10 bins is that the population of every bin is limited by 0.002k. Too many bins would result in large noise in every bin. At the same time, through this process, we have produced a large sample of ellipsoids distorted by the string because it is 100 times more than a single step. Such a large sample would realize ED and EAD of the lensed galaxies with small statistical errors.

\begin{figure*}
\centering
\includegraphics[width=50mm]{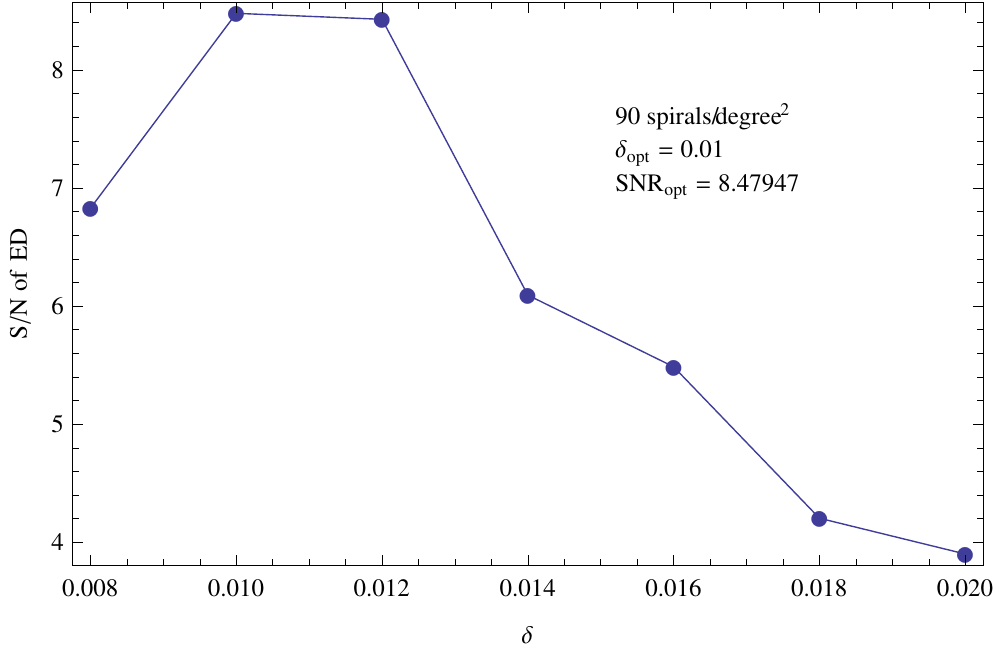}
\includegraphics[width=50mm]{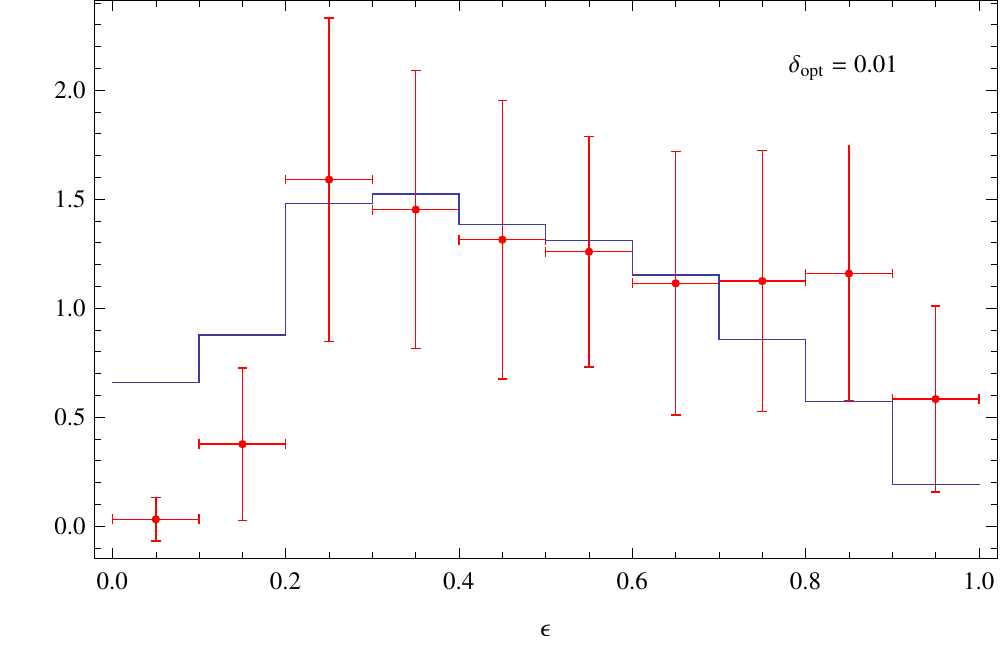}
\includegraphics[width=50mm]{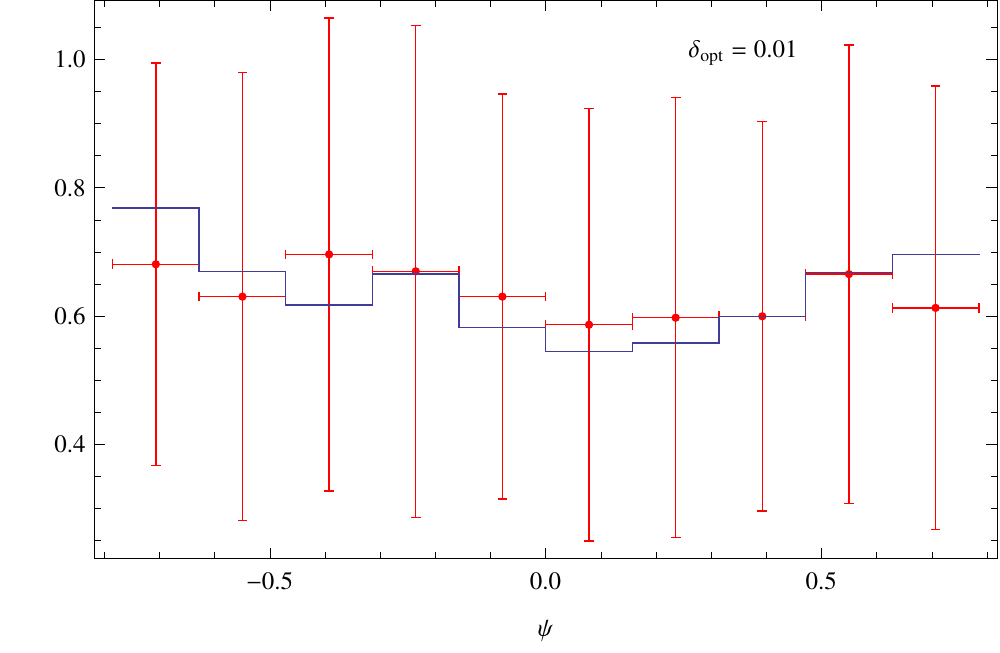}

\includegraphics[width=50mm]{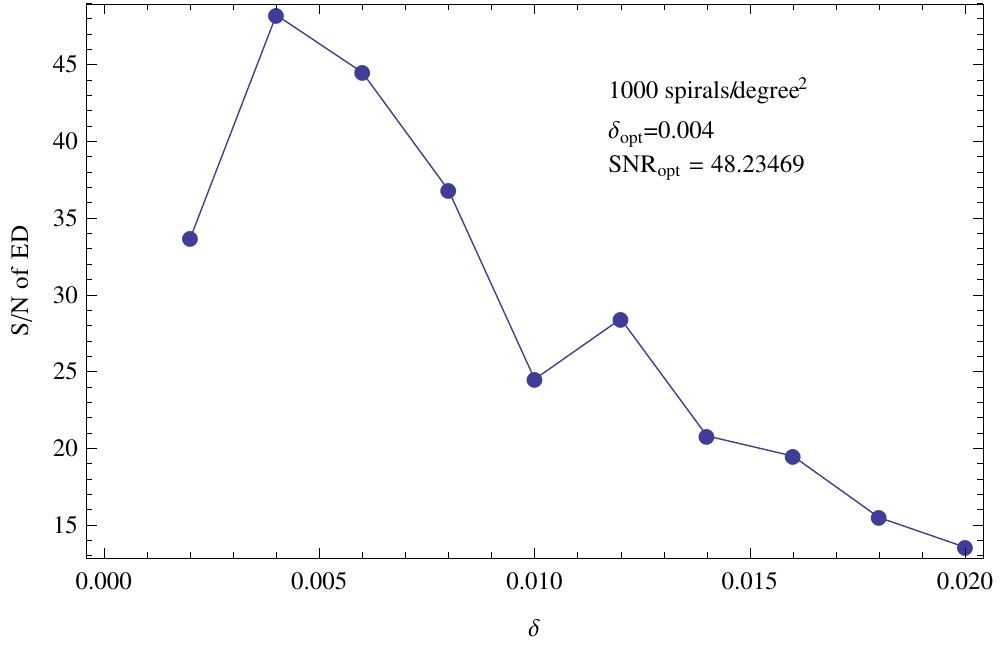}
\includegraphics[width=50mm]{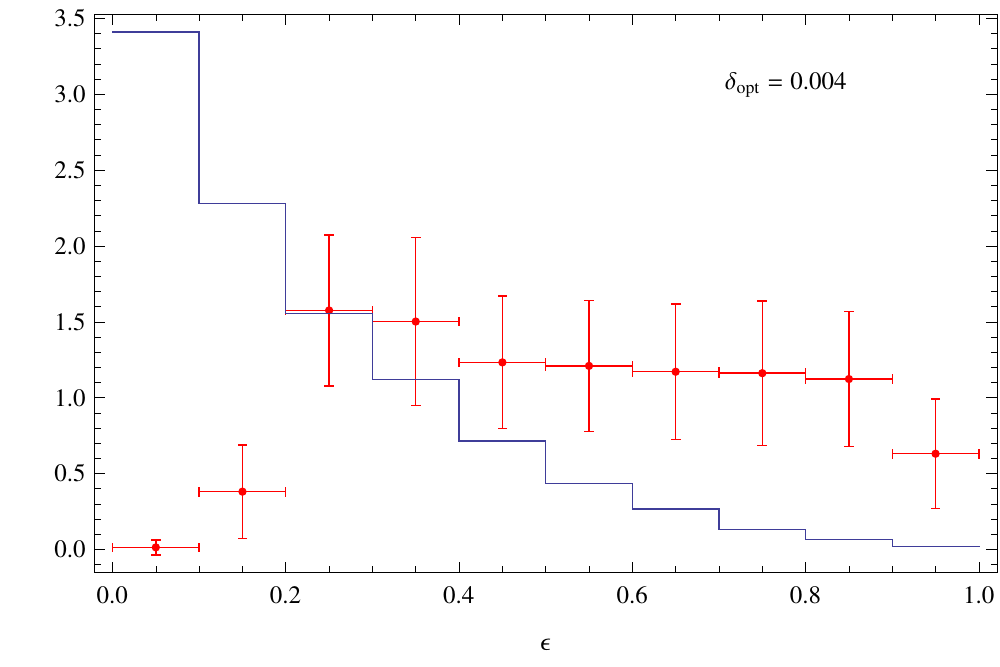}
\includegraphics[width=50mm]{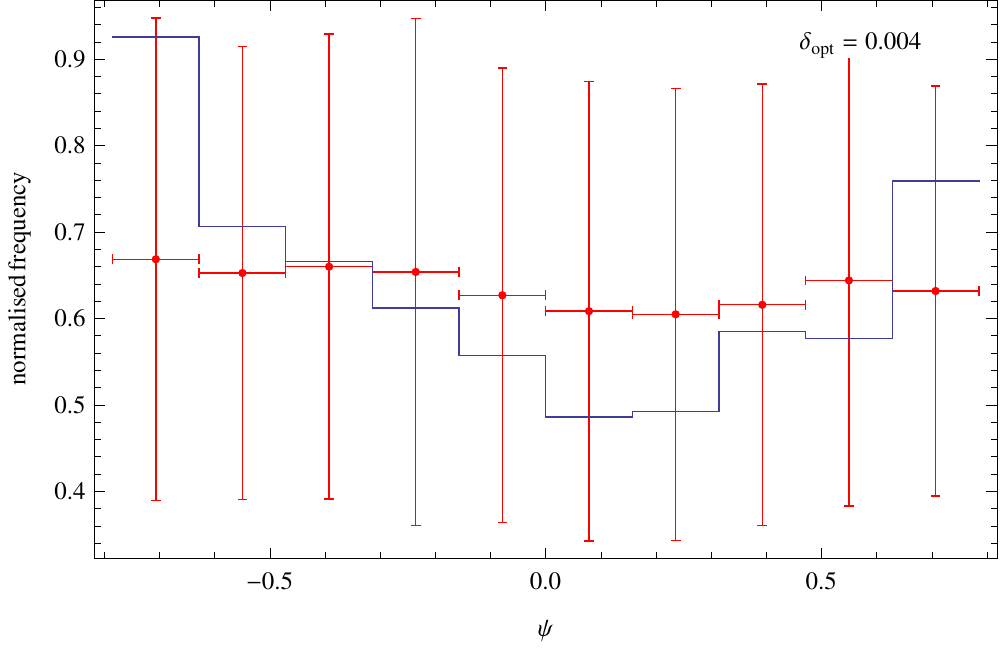}

\includegraphics[width=50mm]{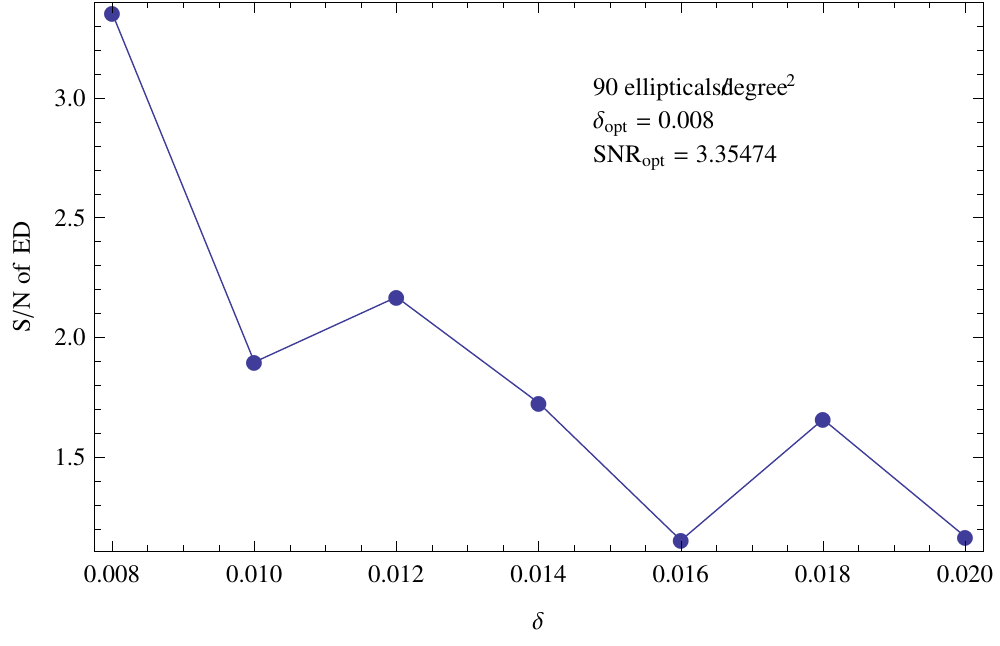}
\includegraphics[width=50mm]{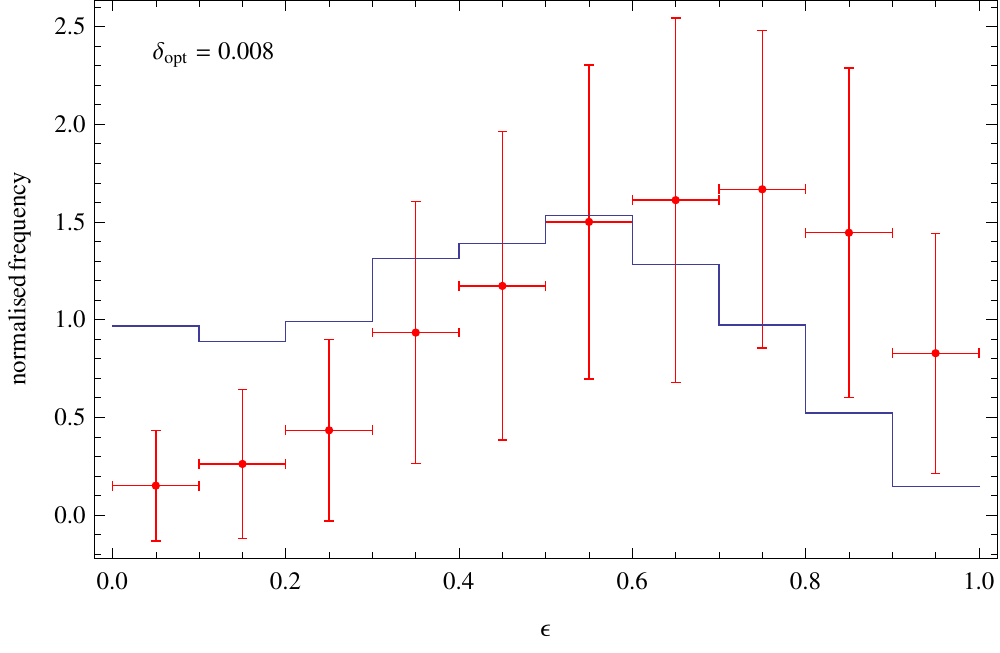}
\includegraphics[width=50mm]{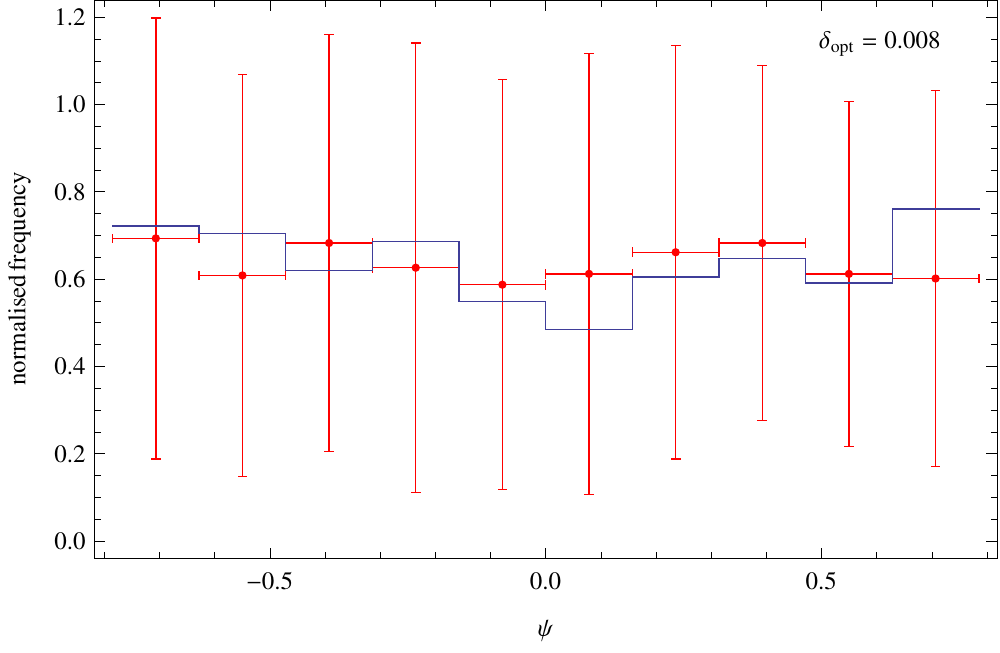}

\includegraphics[width=50mm]{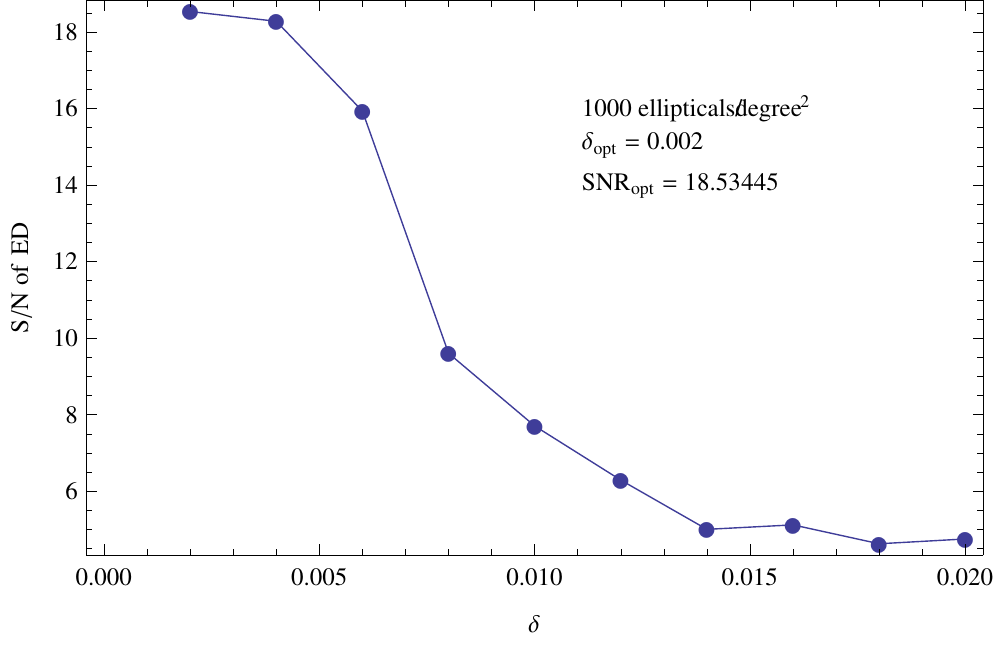}
\includegraphics[width=50mm]{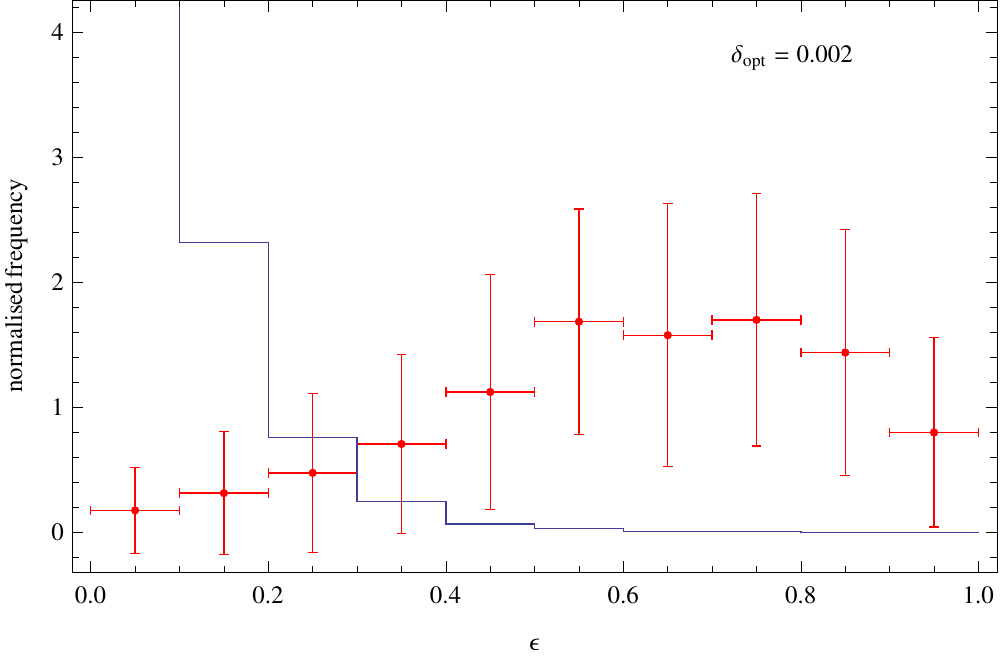}
\includegraphics[width=50mm]{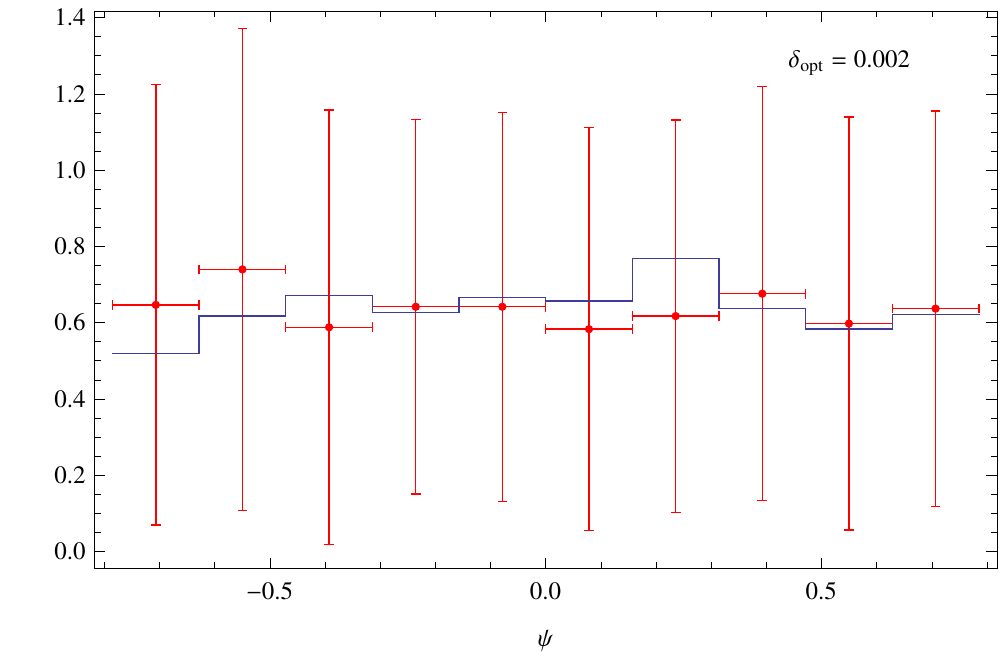}
\caption{Left panels show SNRs in the first bin as functions of $ \delta $ for different surface number densities and different EDs. Middle and right panels show histograms of lensed EDs and EADs with $\delta_{opt} $. Each $\delta_{opt} $ is given by the left corresponding graph. The error bars in the first and the third (the second and the fourth) histograms are generated by choosing the point with its parameters obeying the above distributions about $ n=90~\Delta\theta^2 180^2/\pi^2 $ ($ n=1000~\Delta\theta^2/\pi $) times and repeating this procedure 100 times. The spirals and ellipticals are created through $100\times n$ random choices of $\Delta\theta,\Delta\phi\in[-0.5\delta, 0.5\delta] $ and $ \{\zeta,\xi\} $ according to the ellipticity distributions by Padilla and Strauss \citep{PaSt08}, and randomly choosing the points on an unit sphere which gives the distributions of $ \alpha, \beta,\gamma$}
\label{fig:SNR}
\end{figure*}

Third, subtracting the original ED/EAD from the lensed ED/EAD, we extracted the useful signal. Then, divide the signal by the standard deviation of the original ED/EAD, we got SNR as a function of $ \delta$ (Fig.\ref{fig:SNR}).

 We adopted two different ED models, one of spirals and the other of ellipticals, and computed SNRs of them respectively. Due to the little contribution of fitting errors and systematic errors to the total error in the first bin, only statistical errors would pollute the signal. So we divided the signal by the statistical error to get SNR. After computing 34 SNR graphs (17 for EDs and 17 for EADs), we found that SNR of ED in the first bin, $ \epsilon<0.1 $, is much larger than that of the others. However, in most of the cases, the SNR of EAD is less than 1. Thus EAD could not be used to detect the cosmic string as Dyda and Brandenberger predicted in their paper \cite{DyBr07}. Thus, the SNR of ED in the first bin is chosen to mark the level of feasibility. In addition, we chose two surface number densities, one of which is the density of SDSS main galaxy sample and one is the lower limit of the future LSST's density. The types of galaxies also contribute to the variation of EDs and EADs. Thus, we have four optimal modes corresponding to four distributions: $ 2~ types ~of ~galaxies\times 2~ kinds~ of ~number~ densities $. Each mode require one specific survey. For example, it would be better to divide the sky area into small circles, each of which has an area of $ \delta^2 $. Notice that every circle is supposed to have coinciding area; and every inscribed square of each circle should adjacent to those of surrounding circles. However, if the sample is large enough, the requirement of specific geometry might be weak. We could find, in Fig.\ref{fig:SNR}, four optimal modes, corresponding to $ \delta_{opt} $) for the analysis of the current and future samples.

In Fig.\ref{fig:SNR}, there are four SNRs for four EDs, each of which is corresponding to one specific density and one type of galaxy. It is easy to get the optimal modes of each SNR. The optimal modes for $ 90~spirals/degree^2$, $1000
~spirals/degree^2$, $ 90~ ellipticals/degree^2$, $1000 ~ellipticals/degree^2 $ are $ \Delta\theta_{opt}=$ 0.01,~0.004,~0.008,~0.002. The optimal SNR of spirals with current number density is 8.48 which is larger than that of ellipticals 3.35. With higher density, ellipticals would also be appropriate for abstracting lensing signals because of its high SNR 18.53. The results indicate that the mode of spirals is better to be used as a probe, particularly for current samples with low number density of $90~ spirals/degree^2$. Though spirals suffer more extinction of dust than ellipticals do, the ED of them is still the best probe for cosmic string for its high SNR. Besides, the development of the correction for dust extinction and the division of the sample according to magnitudes would improve the accuracy of this method \citep{PaSt08}. For the future samples with higher number density, it is better to adopt all available modes (with different $ \delta $) rather than only one mode. By the way, the large amount of data would decrease the statistic errors. Under this condition, the systematic error might be larger than the statistic error. The development of this method would require accurate models of EDs for different types of galaxies to decrease systematic errors.

However, the SNR of EAD is too low for usage. Thus we only give the error bar of the EAD with respect to each optimal mode. There is almost no SNR larger than 1. The result indicates that EAD has relatively low SNR and is not so appropriate for detecting cosmic string with current data. The low number density of the sample impedes the development of this method. However, EAD is independent with ED models because the original EAD is always flat. Besides, the release of the future data would make the statistic
 errors lower than systematic errors. A density of $10,000~galaxies/degree^2 $ might be enough for EAD to be used to abstract useful lensing signals. Thus, EAD would even be better than ED in a long run. In addition, for the reason that the future large sample shall weaken the requirement of optimal modes, the usage of different modes simultaneously would of course increase the power of these methods. Nevertheless, whatever modes we use, ED and EAD would be abnormal if the galaxies lied in the direction of the string. The position of these abnormal EDs and the special angle of EADs would reveal the true direction of the string. The reason is that the change of the population in the middle bin of EAD, i.e. $\Psi=0$, is larger than any other bins. This special direction is parallel with x1 axis, which indicate the direction of the cosmic string.

Generally, these methods are different from the one developed by Dyda. Dyda and Brandenberger employed special range of the ellipticity to get obvious EAD \citep{DyBr07}. Here, we have used special position to get abnormal EAD and ED. It is easier to find specific area than to  find specific ellipticity because the improvement of survey area is easier than the correction of ellipticity. Moreover, the future survey will weaken the requirement of survey modes for its lower SNR. For example, the optimal ED mode of ellipticals with high number density is $ \delta_{opt}=0.002$. Under this condition, we can just get a small solid angle of $ \delta^2 180^2/\pi^2\approx 0.013~ deg^2 $. The survey area of LSST will include 30,000 $ deg^2 $ in which a databases of 10 billion galaxies would be produced \citep{ivezic08}. Thus, we can easily get 4000 galaxies in 0.013 $ ~deg^2 $ which would largely decrease the statistic noise and increase SNR greatly. However, if there were no significant variations of ED and EAD, we would put stronger constraints on string qualities, such as $ G (U -T) $. In (14), $\vec{\alpha}$ is in inverse proportion to $ G (U -T) $. That is, there would be two factors contributing to the result that no ED signals were detected. The first one is $ G (U -T)\ll10^{-6} $, and the second one is that the area covered by the survey is not large enough. With larger and deeper sky areas being surveyed, the first factor would be the main factor, and $ G (U -T) $ would get stronger constraints as a result. Moreover, the probe of the cosmic string requires a cone-like and degree by degree survey with enormous galaxy sample in each $ degree^2 $ to increase signal-to-noise ratio (S/N). This kind of survey may be realized by LSST, Dark Energy Survey (DES), etc.

\vspace{4.5mm}

\noindent
\chapter{\large \usefont{T1}{fradmcn}{m}{n}\xbt 4\quad Conclusions}
\vspace{2mm}

\noindent
In this paper, we have studied the weak lensing signal caused by an infinite cosmic string especially by a wiggly string. In the beginning, we utilize a sensitive indicator of gravitational distortion, $ \eta_G $, to replace the non-constant distortion angle. We have studied the straight jet segment and get the probability to detect a cosmic string with current sample and current ability of radio telescope. Moreover, the shape of $ \eta_G $ is different from those of other effects and could be distinguished from various sources. It is impressive that $ \eta_G $ is constant for a straight jet, and $ \eta_G $ looks like a quasi-sinusoidal curve for a spherical lens. This special phenomenon could be used to detect a cosmic string. $ \eta_G $ is in the order of $ 10^{-6} $, but $ \Delta\theta $ and $\Delta\phi$ play an important part to increase the strength of $ \eta_G $. Thus, the position of a jet determines the value range of $ \eta_G $; and the inclined angle of $ \eta_G $ determines the specific value of $\eta_G$. After simple calculation, we can determine the relationship between distribution of jets number density with $ \eta_G $ through simulations. It is the first time that the radio jet is employed as an extended gravitational lensing probe to find a wiggly string. It also supplies an independent method to study the intrinsic properties of the cosmic string. Due to the weak signal of the cosmic string, large scale surveys is required, and more accurate analysis is needed to extract useful signals. It would also be interesting to use Expanded Very Large Array (EVLA) to probe the constant $ \eta_G $. It would identify the special signal yielded by a wiggly string. In addition, we look forward to making use of this tool to detect other different kinds of strings. As the precision of the radio telescope improved in this decade, we would see the appliance of this technique to detect a cosmic string and other objects.

Additionally, ED could also be a sensitive probe of a wiggly string. The
special area where lensed ED and EAD was found would reveal the specific
direction of a wiggly cosmic string. To be specific, a wiggly string would
make galaxies more elliptical when they lie in the direction of the string. When $ \delta=\delta_{opt} $, the SNR of the ED reaches the largest value. This optimal mode is appropriate for us to use the data from SDSS to abstract string's signal. For current survey, spirals have higher SNR than ellipticals do. With the correction of extinction of spirals, we will divide the SDSS sample into small groups according to their magnitudes in order to decrease the systematic errors in the near future. However, limited by the low number density of current sample, EAD would not be appropriate for probing a cosmic string. In the long run, EAD would be better for cosmic string probing because of its independence with ED models. Moreover, future data release would make more geometric modes of the sky available for putting strong constraints on the abundance of string strongly.

To sum up, all of these probes require accurate measurements of a large sample of background objects, and ED/EAD requires special geometry of the sky for current data processing. Because ED is easily affected by other factors, such as dust extinction and reddening, dividing the galaxy sample into sub-samples is necessary for future research on cosmic string detection. It is also important to study
the weak lensing effect of different kinds of cosmic strings and loops. Further, the shear and convergence fields of cosmic strings need to be carefully studied too. The next step is to use the sample of SDSS to get ED and EAD in each small sky area. With more rigorous statistical analyses of various errors, the constraints from the real data would be strong enough to give upper limit of $ G(T-U)/c^2 $.
\vspace{4.5mm}

\normalem
\section*{Acknowledgement}
Thanks to Y.F. Huang and Shi Qi for helpful input. I am also indebted to H.Brandenberger, P.P. Kronberg and the anonymous reviewer for many valuable suggestions. This work was supported by the National Natural Science Foundation of China (Grant No. 10625313) and the National Basic Research Program of China(973 Program, Grant 2009CB824800).
\end{multicols}

\noindent\rule{\textwidth}{1pt}
\begin{multicols}{2}
\bibliographystyle{utphys}
\bibliography{radio_jets_and_galaxies_as_cosmic_string_probes}
\end{multicols}

\end{document}